\documentclass[reprint,amsmath,amsfonts,amssymb,aps,prx,preprintnumbers,superscriptaddress]{revtex4-2}

\usepackage[utf8]{inputenc}
\usepackage[T1]{fontenc}
\usepackage{lmodern}
\usepackage{graphicx}
\usepackage{dcolumn}
\usepackage{amsmath}
\usepackage{bm}
\usepackage{textcomp}
\usepackage{ulem}
\usepackage{ifpdf}
\usepackage[squaren,Gray]{SIunits}
\usepackage{color}
\definecolor{red}{rgb}{1,0,0}
\definecolor{blue}{rgb}{0,0,1}
\definecolor{darkred}{rgb}{0.6,0,0}
\definecolor{darkblue}{rgb}{0,0,.6}
\definecolor{darkgreen}{rgb}{0,0.5,0}

\usepackage{ifpdf}
\ifpdf
\usepackage{epstopdf}
\usepackage[pdftex,unicode,pdfstartview={FitH},pdfborder={0 0 0}]{hyperref}
\usepackage{hypcap}
\else
\usepackage[hypertex]{hyperref}
\fi
\hypersetup{
    bookmarksnumbered = true,
    colorlinks = true, linkcolor = black,
    citecolor = black, filecolor = black,
    menucolor = black, urlcolor = black
}

\newcommand{\RX}{$R_h^X\phantom{.}$}
\newcommand{\RM}{$R_h^M\phantom{.}$}
\newcolumntype{R}{>{$\displaystyle}r<{$}}
\newcolumntype{C}{>{$\displaystyle}c<{$}}

\begin{document}
    
    \title{Control of atomic reconstruction and quasi-1D excitons in strain-engineered \\ moir\'e heterostructures}
    
     \author{Shen Zhao}
    \affiliation{Fakult\"at f\"ur Physik, Munich Quantum Center, and Center for NanoScience (CeNS), Ludwig-Maximilians-Universit\"at M\"unchen, Geschwister-Scholl-Platz 1, 80539 M\"unchen, Germany}

    \author{Zhijie Li}
    \affiliation{Fakult\"at f\"ur Physik, Munich Quantum Center, and Center for NanoScience (CeNS), Ludwig-Maximilians-Universit\"at M\"unchen, Geschwister-Scholl-Platz 1, 80539 M\"unchen, Germany}
    \affiliation{Institute of Physics, Carl von Ossietzky University, 26129 Oldenburg, Germany}
    
    \author{Zakhar A. Iakovlev}
    \affiliation{Ioffe Institute, 194021 Saint Petersburg, Russian Federation}

	\author{Peirui Ji}
    \affiliation{State Key Laboratory for Manufacturing Systems Engineering, Xi'an Jiaotong University, Xi'an 710049, China}
    
    \author{Fanrong Lin}
    \affiliation{1st Physical Institute, Faculty of Physics, University of G\"ottingen, Friedrich-Hund-Platz 1, G\"ottingen 37077, Germany}  

	\author{Xin Huang}
	\affiliation{Key Laboratory of Nano-devices and Applications, Suzhou Institute of Nano-Tech and Nano-Bionics, Chinese Academy of Sciences (CAS), Suzhou 215123, China}

    \author{Kenji Watanabe}
    \affiliation{Research Center for Electronic and Optical Materials, National Institute for Materials Science, 1-1 Namiki, Tsukuba 305-0044, Japan}
    
    \author{Takashi Taniguchi}
    \affiliation{Research Center for Materials Nanoarchitectonics, National Institute for Materials Science, 1-1 Namiki, Tsukuba 305-0044, Japan}

    \author{Mikhail M. Glazov}
    \affiliation{Ioffe Institute, 194021 Saint Petersburg, Russian Federation}

\author{Anvar~S.~Baimuratov}
    \affiliation{Fakult\"at f\"ur Physik, Munich Quantum Center, and Center for NanoScience (CeNS), Ludwig-Maximilians-Universit\"at M\"unchen, Geschwister-Scholl-Platz 1, 80539 M\"unchen, Germany}
    \affiliation{Center for Engineering Physics, Skolkovo Institute of Science and Technology, Bolshoy Boulevard 30, building 1, Moscow 121205, Russia}

\author{Alexander H{\"o}gele}
    \affiliation{Fakult\"at f\"ur Physik, Munich Quantum Center, and Center for NanoScience (CeNS), Ludwig-Maximilians-Universit\"at M\"unchen, Geschwister-Scholl-Platz 1, 80539 M\"unchen, Germany}
    \affiliation{Munich Center for Quantum Science and Technology (MCQST), Schellingstra\ss{}e 4, 80799 M\"unchen, Germany}
    \date{\today}
\begin{abstract}

In two-dimensional nearly commensurate heterostructures, strain plays a critical role in shaping electronic behavior. While previous studies have focused on random strain introduced during fabrication, achieving controlled structural design has remained challenging. Here, we demonstrate the deterministic creation of one-dimensional arrays from initially zero-dimensional triangular moir\'e patterns in MoSe$_2$-WSe$_2$ heterobilayers. This transformation, driven by the interplay of uniaxial strain and atomic reconstruction, results in one-dimensional confinement of interlayer excitons within domain walls, exhibiting near-unity linearly polarized emission related to the confinement-induced symmetry breaking. The width of the domain walls--and consequently the degree of exciton confinement--can be precisely tuned by the interlayer twist angle. By applying out-of-plane electric field, the confined excitons exhibit energy shifts exceeding 100~meV and changes in the fine-structure splitting by up to a factor of two. Our work demonstrates the potential of strain engineering for constructing designer moir\'e systems with programmable quantum properties, paving the way for future optoelectronic applications.
\end{abstract}

\maketitle

Moir\'e superlattices based on two-dimensional (2D) van der Waals materials provide a powerful platform for exploring novel quantum phenomena with exceptional tunability through structural and electrical control~\cite{Andrei2021}. Since the discovery of superconductivity~\cite{Cao2018}, graphene-based moir\'e systems have been found to exhibit a variety of exotic states, including correlated insulators~\cite{Cao2018b}, Chern ferromagnets~\cite{Chen2020}, and fractional metals~\cite{Zhou2021}. Likewise, moir\'e superlattices of semiconducting transition metal dichalcogenides (TMDs) have emerged as a complementary system, enabling the study of bosonic and topologically non-trivial quantum phases with optical accessibility~\cite{Mak2022}, such as excitonic insulators~\cite{Gu2022}, exciton density waves~\cite{Zeng2023}, and fractional quantum anomalous Hall effects~\cite{Cai2023}. 

The electronic properties of bilayer moir\'e systems depend crucially on the periodicity, symmetry, topology and dimensionality of the underlying microscopic structure~\cite{Nuckolls2024}. Notably, compared to individual layers, the effects of strain on bilayers are particularly important and scale as the inverse of interlayer twist angle $\theta$. As a result, strain can significantly modify the original canonical moir\'e superlattice~\cite{Escudero2024,Koegl2023}. Meanwhile, to minimize the stacking energy, the superlattice tends to spontaneously rearrange into a domain-wall pattern, particularly at small twist angles~\cite{Yoo2019,Weston2020}. This atomic reconstruction phenomenon, along with random strain generated during fabrication, have been identified as major factors contributing to complex spatial variations in realistic moir\'e samples~\cite{Kazmierczak2021,Halbertal2021,Jong2022,Halbertal2022,VanWinkle2023}, which in turn dramatically affect their transport and optical behaviors~\cite{Uri2020,Zhao2023}. Better control over these two structural factors offers great potential for engineering customized moir\'e superlattices.


\begin{figure*}[t!]
    \includegraphics[scale=0.98]{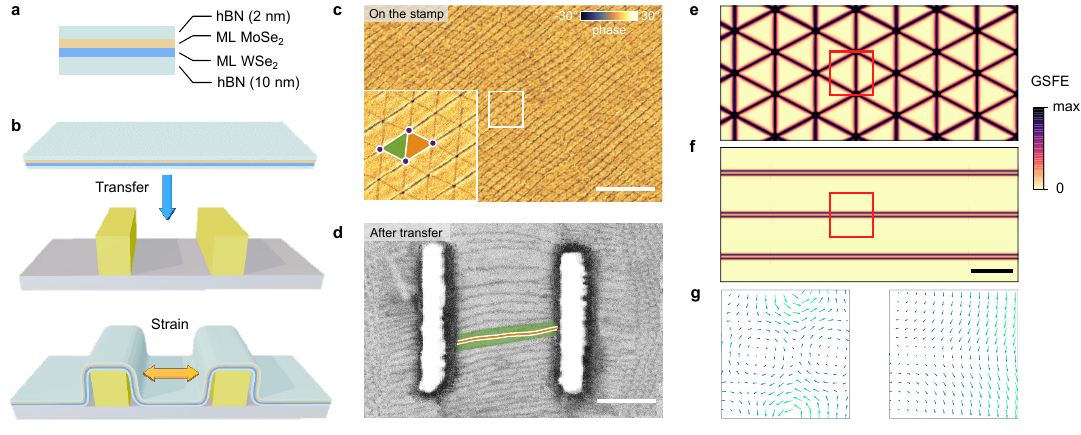}
    \caption{\textbf{Deterministic creation of 1D arrays in MoSe$_2$-WSe$_2$ heterobilayers via strain-engineering.} \textbf{a}, Schematic cross-section of a representative MoSe$_2$-WSe$_2$ heterobilayer sample encapsulated in thin hBN layers, with the thickness of the top and bottom hBN given in parentheses. \textbf{b}, Illustration of the experimental method to locally induce uniaxial strain on the heterobilayer. After assembly using a polymer stamp, the flat sample is transferred onto a patterned substrate featuring parallel elongated mesas. Due to strong adhesion, the sample conforms to the mesa contours, resulting in uniaxial strain between the mesas. \textbf{c}, TFM phase image of the heterobilayer before transfer, showing a homogeneous region that features a 0D periodic lattice. The inset highlights the alternating triangular pattern of the domains, which correspond to the two energetically favorable stackings \RM and \RX (green and orange, respectively). \textbf{d}, SEM image of the same region as in \textbf{c} after transfer, revealing the formation of a 1D arrays consisting of elongated \RM domains (green) separated by narrow domain walls of \RX stacking (orange). \textbf{e} and \textbf{f}, Calculated generalized stacking fault energy maps for a triangular moir\'e pattern with $\theta = 0.4^\circ$ and the corresponding 1D pattern. \textbf{g},~Left (right) panel: deformation field in the area indicated by the red square in panel \textbf{e} (\textbf{f}). Scale bars: 500~nm in \textbf{c} and \textbf{d}, 50~nm in \textbf{f}.}
    
    \label{fig1}
\end{figure*}

In this work, we exploit the interplay between uniaxial tensile strain and atomic reconstruction to achieve programmable generation of one-dimensional (1D) arrays in MoSe$_2$-WSe$_2$ moir\'e heterobilayers. Combining generalized stacking fault energy simulations with electron channeling contrast imaging, we reveal that these 1D arrays consist of elongated domains stabilized in the lowest-energy stacking configuration and separated by parallel domain walls composed of metastable stacking configurations transitioning from triangular to linear geometries. Remarkably, we observe strongly linearly polarized emission from interlayer excitons confined within the domain walls and related to the confinement-induced symmetry breaking. By correlating nanoscale structural mapping with cryogenic micro-photoluminescence (PL) spectroscopy, we establish a direct relationship between the domain wall width -- controlled via interlayer twist angle -- and the quantum confinement strength of interlayer excitons. Furthermore, exploiting the permanent out-of-plane dipole moment of excitons, we realize electrically tunable energy shifts exceeding $100$~meV for 1D-confined interlayer excitons while preserving their high degree of linear polarization. This work establishes strain-engineered moir\'e heterostructures as a versatile platform for creating tunable 1D quantum systems, thus advancing the development of on-demand quantum material design.

Figure~\ref{fig1}a illustrates schematically the cross-sectional structure of a representative sample fabricated for stain engineering. By a layer-by-layer all-dry stamping technique (Methods), MoSe$_2$ and WSe$_2$ monolayers synthesized by chemical vapour deposition were aligned to around $0^\circ$ (R-type) and encapsulated by hexagonal boron nitride (hBN) layers. The 2-nm-thick top hBN layer enables direct visualization of moir\'e superlattice via nano-imaging, while the $10$-nm-thick bottom hBN layer ensures both optical quality~\cite{Cadiz2017} and mechanical coupling to the patterned substrate. As illustrated in Fig.~\ref{fig1}b, to apply local strain to the heterobilayer, we designed a substrate patterned with parallel pairs of elongated mesas (see Supplementary Fig.~3 for the entire pattern of 3 $\times$ 3 pairs) and transferred the sample onto it. These mesas, evaporated from gold and chromium, have a footprint of 1.5~\micro\meter~$\times$~0.2~\micro\meter~and a total height of 160~nm. For each pair, the distance between the two mesas was designed to be 1~\micro\meter. After the transfer, the originally flat sample conforms to the contours of the patterned substrate due to strong van der Waals forces, causing the region between the mesa pair to stretch along the direction perpendicular to the mesas. As a result, local uniaxial tensile strain is applied onto the heterobilayer, as demonstrated previously in a similar approach for bulk MoS$_2$~\cite{Dong2023}.

\begin{figure*}[t!]
    \includegraphics[scale=0.98]{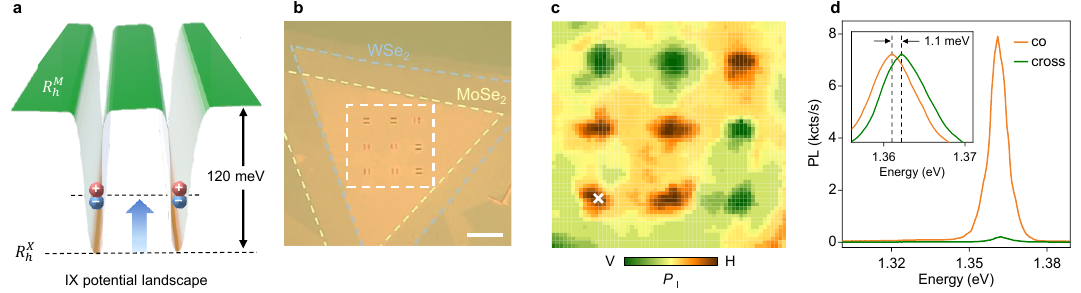}
    \caption{\textbf{Confinement of interlayer excitons within the domain walls of 1D arrays.} \textbf{a}, Schematic of the interlayer exciton (IX) potential landscape in a 1D array region. The narrow width of \RX domain walls and the high energy barrier from adjacent \RM domains confine interlayer excitons within the domain walls, resulting in a blue-shift of their energy compared to the free exciton energy of \RX stacking (arrow). \textbf{b}, Optical micrograph of the sample after transfer onto a patterned substrate with nine pairs of horizontally or vertically oriented mesa pairs. The dashed yellow and blue lines indicate the locations of monolayer MoSe$_2$ and WSe$_2$, respectively. The scale bar is 5~\micro\meter. \textbf{c}, 2D spatial map of linear polarization ($P_\mathrm{l}$) of interlayer exciton emission within the area marked by the white box in \textbf{b}. Orange and green, respectively, indicate horizontal (H) and vertical (V) polarization, defined by the orientation of the micrograph in \textbf{b}. The hot spots align with the positions of mesa pairs, with polarization matching the domain wall axis orthogonal to the elongated mesas. \textbf{d}, Linear polarization-resolved PL spectra of 1D-confined interlayer excitons, contrasting co- and cross-polarized components relative to the domain wall axis, acquired at the location marked by the cross symbol in \textbf{c}. The inset shows normalized spectra highlighting the energy splitting of linearly polarized interlayer exciton states.}
    \label{fig2}
\end{figure*}

To reveal the effect of strain on the moir\'e structure, we performed real-space nano-imaging on the heterobilayer before and after transfer. Initially, when the heterobilayer was on the stamp prior to capping with the bottom hBN layer, we mapped its structure using torsional force microscopy (TFM). This atomic force microscopy-based technique offers a non-destructive method to visualize moir\'e superlattices with lateral resolution below 1~nm~\cite{Pendharkar2024}. Figure~\ref{fig1}c presents a typical TFM phase image recorded in the core of the heterobilayer. We observe an uniform zero-dimensional (0D) lattice formed by periodic triangular domains approximately 60~nm in size, consistent with theoretical predictions of lattice reconstruction in R-type heterobilayers~\cite{Carr2018,Enaldiev2020}. Note that the 0D lattice actually consists of two alternating energetically favored stackings \RM and \RX (highlighted in green and orange)~\cite{Weston2020,Rosenberger2020}, however, TFM can only distinguish the boundaries between them, appearing as thin lines of darker contrast. After characterization of the entire heterobilayer with TFM, we transferred the sample by aligning the homogeneous 0D regions with the mesa pairs to avoid random strain, which can also change the reconstruction pattern~\cite{Zhao2023}.

Following the transfer process, we carried out electron channeling contrast imaging in scanning electron microscopy (SEM) to visualize the heterobilayer. In contrast to TFM, this technique differentiates the two optimal stackings \RM and \RX according to their distinct electron channeling conditions~\cite{Andersen2021,Rupp2023}. Figure~\ref{fig1}d shows a SEM image of the same region as Fig.~\ref{fig1}c for comparison. We find that the applied uniaxial strain causes the triangular domains of \RM stacking to expand and merge into an array of elongated domains (example marked in green), bridging the two mesas and maintaining a width of $\sim60$~nm. Meanwhile, the \RX stacking collapsed into few-nanometer-wide double domain walls that separate the \RM domains on both sides (marked in orange). This local transformation from 0D lattice into 1D array also occurred in other mesa pairs of the same sample as well as other samples (see Supplementary Fig.~4 for more examples).  

To explain the mechanism of strain-induced 1D array formation, we adopt the theoretical model described in Refs.~\cite{Enaldiev2020,Zhao2023}, explicitly including the effect of uniaxial strain. It is established that the formation of a 1D array consisting of elongated domain walls can become energetically more favorable as compared to a triangular network with nodal domain walls~\cite{Alden2013,Lebedeva2019May,Enaldiev2024Apr}. In our case, we demonstrate that uniaxial strain decreases the symmetry and effectively modifies the boundary conditions in simulations, leading to a periodic 1D array of energetically optimal configuration, as indicated by generalized stacking fault energy (GSFE) or adhesion energy maps~\cite{Carr2018,Enaldiev2020} in Figs.~\ref{fig1}e and \ref{fig1}f. The simulations show the formation of elongated domains of \RM stacking and double domain walls with the core exhibiting \RX stacking, consistent with the experimental results. Figure~\ref{fig1}g illustrates the corresponding displacement fields for these two configurations. Further details of the theoretical model and simulations are provided in Supplementary Note~1.

\begin{figure*}[t!]
    \includegraphics[scale=0.97]{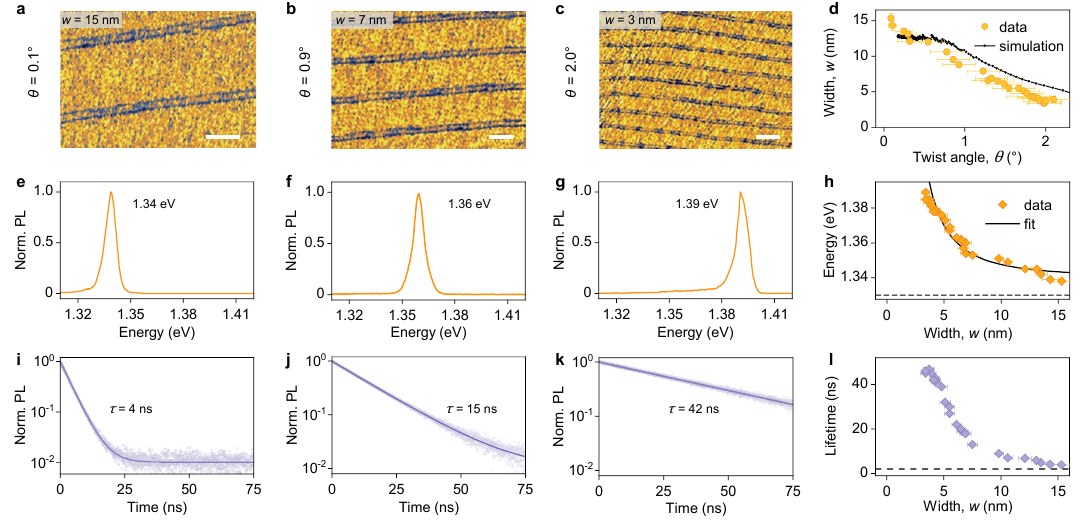}
    \caption{\textbf{1D-confined interlayer excitons in domain walls of varying widths.} \textbf{a,b,c}, TFM phase images of three 1D arrays with domain wall width $w$ of 15~nm (\textbf{a}), 7~nm (\textbf{b}), and 3~nm (\textbf{c}). The corresponding local twist angles $\theta$ before strain are shown on the left (calculated from the average triangular domain size, see Methods for details). The scale bars are 50~nm, 20~nm and 15~nm, respectively. \textbf{d}, Plot of $w$ as a function of $\theta$, summarizing all strain-induced 1D arrays in three samples. Vertical and horizontal error bars represent $95\%$ confidence intervals for $w$ and $\theta$, respectively. Black dots and line represent the simulated domain wall width based on the atomic reconstruction model (see Supplementary Note~3). \textbf{e-l}, Correlation of 1D-confined interlayer excitons with domain wall width $w$. Normalized PL spectra (\textbf{e,f,g}) and PL decay (\textbf{i,j,k}) acquired from the 1D arrays shown in \textbf{a-c}. The energy position of peak maximum obtained from Lorentzian fitting is indicated in each PL spectrum. The PL decay traces are fitted with single exponential functions (lines) with lifetimes indicated on the side. \textbf{h,l}, Dependence of interlayer exciton energy (\textbf{h}) and lifetime (\textbf{l}) on $w$. Horizontal and vertical error bars depict $95\%$ confidence intervals in $w$ and uncertainties from the least-square fits, respectively. Dashed lines in \textbf{h} and \textbf{l} denote the energy and lifetime of free interlayer excitons in \RX stacking, determined from a micron-sized 2D domain (Supplementary Fig.~5). Solid line in \textbf{h} represents the fit based on the infinite potential well model.}
    \label{fig3}
\end{figure*}

We now explore the optical properties of the strain-induced 1D arrays. \textit{ab initio} calculations have estimated the interlayer energies of \RX and \RM stackings as 1.33 and 1.45~eV, respectively~\cite{Forg2021}. As schematically illustrated in Fig.~\ref{fig2}a, the significant energy difference of $\sim120$~meV strongly modulate the potential landscape in the 1D array, resulting in efficient trapping of interlayer excitons within the domain wall of \RX stacking. Compared to free excitons in \RX stacking, the 1D-confined excitons are supposed to switch their optical selection rule from circular to linear polarization along the domain wall axis and have a sizable blue-shift in energy~\cite{Zhao2023}. To verify this assumption, we perform linear polarization resolved PL measurements at $\sim4$~K. Figure~\ref{fig2}b and c, respectively, show an optical micrograph of the sample after applying strain and the map of the degree of linear polarization ($P_\mathrm{l}$) in the region of mesa pairs. Here, $P_\mathrm{l}$ is defined as $P_\mathrm{l} = (I_\mathrm{H}-I_\mathrm{V})/(I_\mathrm{H}+I_\mathrm{V})$, where $I_\mathrm{H}$ ($I_\mathrm{V}$) represents the integrated PL intensity detected under the polarization along the horizontal (vertical) direction of the optical micrograph in Fig.~\ref{fig2}b. We observe that all nine strain-induced 1D arrays exhibit near-perfect linear polarization ($|P_\mathrm{l}| > 0.9$) with the angle aligned along the domain wall axis and perpendicular to the corresponding mesa orientation. The determined polarization angle strongly contrasts with the results from randomly strained MoSe$_2$-WSe$_2$ heterobilayers, where the occurrence of linearly polarized interlayer exciton PL is arbitrary and the $|P_\mathrm{l}|$ values are relatively low~\cite{Alexeev2020,Bai2020}.

Figure~\ref{fig2}d presents linear polarization-resolved PL spectra of 1D-confined interlayer excitons under 1~\micro\watt~excitation, with an emission peak of 7~meV linewidth centered at $\sim$1.36~eV. The peak energy exhibits a pronounced 30~meV blue shift compared to free interlayer excitons in 2D \RX domains (Supplementary Fig.~5), consistent with the size-quantization effect in the potential landscape described in Fig.~\ref{fig2}a. Notably, the co- and cross-polarized components display a non-negligible splitting of 1.1~meV (inset of Fig.~\ref{fig2}d), contrasting with the valley-degenerate circular polarization of free interlayer excitons. With increased excitation power, a secondary peak emerges 20~meV above the primary emission, with identical polarization characteristics and energy splitting (Supplementary Fig.~6). This power-activated higher-energy feature is reminiscent of excited excitonic states observed in the systems with strong confinement~\cite{Thureja2022}. 

To gain insight into the above behaviors of 1D-confined interlayer excitons, we develop a microscopic theory that takes into account long-range exchange interaction of the electron and hole that constitute the exciton. We model the 1D confinement of exciton center of mass by a rectangular well with infinite barriers (see Supplementary Note~2 for details) which also provides a reasonable agreement for the size-quantization energy, see below. Microscopic calculation based on Refs.~\cite{glazov2014exciton,prazdnichnykh2020control} shows that the splitting of the bright exciton radiative doublet into linearly polarized components in such structures amounts to $|\Delta E| = \pi \hbar \Gamma_0/(\varepsilon_{\rm eff} q w)$, where $\Gamma_0$ is the exciton radiative decay rate of ``bulk'' 2D interlayer exciton and $q = \omega/c$. The estimates show that $|\Delta E| \sim 1\ldots 2$~meV can be easily achieved for the parameters of our structure. Moreover, the order of polarized states observed is consistent with the model: the lower energy sublevel is polarized along the 1D domain walls. At the experimental temperature of 4~K, the higher sublevel cannot be thermally populated due to the substantial energy splitting, which explains the observed near-unity linearly polarized PL along the ensemble of domain walls. Additionally, electrodynamic effects arising from the proximity of gold to the domain walls may further enhance the degree of linear polarization~\cite{Ils1995}.


Harnessing the sub-nanometer resolution of TMF imaging, we are able to precisely measure the \RX-stacked domain wall width $w$. Figure~\ref{fig3}a-c displays representative TFM phase images of 1D arrays from different mesa pairs, with $w$ of 15, 7, and 3~nm, respectively. Interestingly, we find an inverse relationship between the pre-transfer twist angle $\theta$ and post-strain domain wall width: smaller $\theta$ values yield broader domain walls. This trend is further validated by statistical mapping of $w$ versus $\theta$ for all strain-engineered 1D arrays (Fig.~\ref{fig3}d). We qualitatively reproduce this dependence with our theoretical model (black dots, see Supplementary Note~1 for details). However, the numerical simulations overestimate the domain wall widths at relatively large angles. We attribute this discrepancy to an underestimated GSFE for intermediate stacking configurations in the simulations, resulting in enlarged domain wall areas~\cite{Halbertal2021}. Notably, the angle-dependence of the double domain walls provides a more flexible parameter for tuning their width, in contrast to single domain walls that separate domains with different stacking and whose widths are predominantly governed by shear-to-tensile strain ratios~\cite{Alden2013}.

In the following, we correlate the double domain wall width and the optical response of 1D-confined interlayer excitons. Figures~\ref{fig3}e-g and \ref{fig3}i-k present PL spectra and time-resolved PL dynamics acquired from the regions in Figs.~\ref{fig3}a-c. The widest domain wall ($w = 15$~nm) exhibits a PL peak at 1.34~eV (Fig.~\ref{fig3}e), blue-shifted by 10~meV relative to the free interlayer excitons. Their monoexponential PL decay yields a 4~ns lifetime (Fig.~\ref{fig3}f), slightly longer than 2~ns characteristic of unconfined excitons (Supplementary Fig.~4c). As summarized in Figs.~\ref{fig3}h and l, the progressive narrowing of $w$ induces gradual spectral blue-shift and lifetime extension, reaching maxima of $1.39$~eV ($60$~meV shift) and $42$~ns for $w = 3$~nm (as exampled in Figs.~\ref{fig3}g and k). The trends in PL energy are quantitatively captured by the infinite potential well model (solid line in Fig.~\ref{fig3}h), for which we employ a simple fit to the confinement energy given by $E(w) = E_X + \hbar^2 \pi^2/(2 M w^2)$, with $E_X = 1.34$~eV and an exciton mass $M = 0.5\,m_0$ (where $m_0$ denotes the free electron mass). For PL lifetimes, however, the trends cannot be adequately described by a simple inverse proportional relationship with $w$. A more complex model that incorporates the influence of dielectric environments may be necessary~\cite{Glazov2011}, which we defer to future studies. Nonetheless, the wide-range scaling of energy and lifetime with $w$ suggests twist angle as an efficient tuning knob for tailoring the properties of 1D-confined interlayer excitons.

\begin{figure}[t!]
    \includegraphics[scale=0.98]{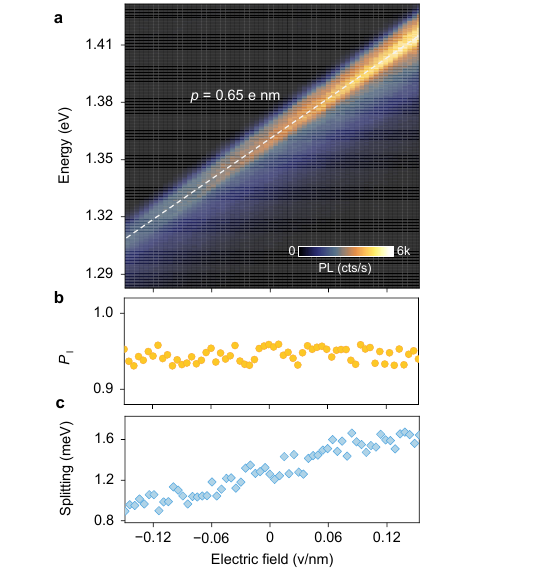}
     \caption{\textbf{Electrostatic control of 1D-confined interlayer excitons.} \textbf{a,b,c}, Evolution of the PL spectra (\textbf{a}), the degree of linear polarization $P_\mathrm{l}$ (\textbf{b}) and the energy splitting between linearly polarized states (\textbf{c}) as a function of out-of-plane electric field. Data recorded from a 1D array in a dual-gated sample, which includes additional top, bottom, and contact electrodes made from few-layer graphene (see Methods).}
    \label{fig4}
\end{figure}

Due to the permanent out-of-plane electric dipole, external electric field can efficiently shift the energy of interlayer excitons through the quantum-confined Stark effect~\cite{Jauregui2019}. To examine this tunability for the 1D-confined excitons, a dual-gated sample containing strain-induced 1D arrays was fabricated (see Methods for details). Figure~\ref{fig4}a displays the linear energy shift of the 1D-confined interlayer excitons with electric field, yielding a dipole moment of 0.65~$e$~nm according to the linear slope, in good agreement with \textit{ab initio} estimation of layer separation for \RX stacking~\cite{Gillen2018}. Remarkably, over an energy range of $\sim$100~meV (from 1.31 to 1.41~eV), the 1D-confined excitons preserve near perfect linear polarization (Fig.~\ref{fig4}b), achieving unprecedented in-situ tunability while maintaining anisotropic emission -- a combination unmatched in other 1D quantum systems~\cite{Glueckert2018}. In addition to the energy shifts, the PL intensity also changes with the electric field, since the electron and hole wavefunctions and thus the optical transition matrix elements of the 1D-confined interlayer excitons change, resulting in an increase in $\Gamma_0$~\cite{Barre2022,FariaJunior2023}. Consistently, the energy splitting between the two linearly polarized states increases from 0.8~meV to 1.6~meV over the electric field range (Fig.~\ref{fig4}c), highlighting again the high tunablility of the system. It has been recently reported that under high fields (above 1~V/nm) the domain walls in twisted homobilayers are possible to move due to ferroelectric effects~\cite{Weston2022}. The related field-driven domain wall dynamics in heterobilayers as well as their implications for interlayer excitons cannot be excluded and remain to be addressed in future work.

In summary, we have demonstrated deterministic control of reconstruction patterns in MoSe$_2$-WSe$_2$ heterobilayers through local uniaxial strain. The resulted 1D arrays feature strong 1D confinement of interlayer excitons, exhibiting near perfect linearly polarized emission ($P_\mathrm{l}>0.9$). The emission energy of these 1D-confined excitons can be tuned by up to 100~meV via twist angle control or by applying an out-of-plane electric field. We anticipate that our strain-engineering approach can be extended to other 2D heterostructures, advancing research into various 1D-related topological~\cite{Kitaev2001,Tong2017,Barrier2024} and strongly correlated phenomena~\cite{Tsunetsugu1997,Li2024}.

\section*{Methods}
\textbf{Sample fabrication:} The mesa patterns (length: 1.5~\micro\meter, width: 0.2~\micro\meter, height: 150~nm and separation distance for each pair: 1~\micro\meter) were fabricated via electron-beam lithography (Raith eLINE) using poly(methyl methacrylate)(PMMA) resist (molecular weight: 950k). First, the resist was spin-coated (thickness: $\sim$450~nm) onto an Si/SiO$_2$ substrate (SiO$_2$ thickness: 285~nm) and exposed under the electron beam with a dose of 500~\micro C/cm$^2$. After development of the resist, 10~nm Cr and 150~nm Au were deposited by an electron-beam evaporator (Von Ardenne, LS 320) followed by lift-off. Finally, a mild sonication or mechanical swab was employed to reduce the roughness of mesas caused by the sidewall effect.  

The hBN encapsulated MoSe$_2$-WSe$_2$ bilayer samples were fabricated by a dry-transfer method~\cite{Pizzocchero2016}. First, triangular MoSe$_2$ and WSe$_2$ monolayers were grown by in-house chemical vapor deposition synthesis and thin hBN were exfoliated from bulk crystals (NIMS). After the determination of thickness by atomic force microscopy, the flakes were successively picked up using Poly(Bisphenol A carbonate) (PC) film stamps in the following order: top hBN layer with $\sim$2~nm thickness, MoSe$_2$ monolayer, WSe$_2$ monolayer and bottom hBN layer with $10-20$~nm thickness. In contrast to the top and bottom hBN, which were directly picked-up, for MoSe$_2$ and WSe$_2$ monolayers it was done in two steps to avoid tearing: (i) Release the already picked-up layers onto the target TMD monolayer, followed by four hours of annealing at 300~°C in a vacuum chamber with a pressure of $10^{-10}$~mbar and (ii) re-pick up the layers together with the TMD monolayer. The pick-up temperatures for hBN, MoSe$_2$ and WSe$_2$ monolayers were around 50~°C, 100~°C and 130~°C, respectively. The two TMD monolayer triangles were aligned in parallel to have a global twist angle close to 0°. Before capping the bottom hBN layer, the MoSe$_2$-WSe$_2$ bilayers were characterized by torsional force microscopy to identify homogeneous regions of reconstructed 0D domains. The local twist angle $\theta$ of these regions was deduced according to the relation $L=a_\text{Mo}/\sqrt{\theta^2+\delta^2}$, where $L$ is the measured period of domain pattern, $\delta = 1- a_\text{W}/a_\text{Mo}$ is the lattice mismatch, $a_\text{W}=0.3282$~nm and $a_\text{Mo}=0.3288$ are the lattice constant of MoSe$_2$ and WSe$_2$, respectively~\cite{AlHilli1972}. Once the entire sample was assembled, it was released onto the patterned substrate in a way that the mesa pairs were covered by the reconstructed 0D domain regions. In the last step, the sample was annealed in $10^{-10}$~mbar for 12~h at a temperature of 200~°C to remove polymer residuals and improve interlayer coupling~\cite{Alexeev2017}.

The sample with additional three few-layer graphene electrodes (contact, top gate and back gate) was fabricated with the same layer-by-layer method as the other samples. The few-layer graphene flakes (thickness of $\sim$2~nm) were exfoliated from bulk graphite (HQ graphene). The top and bottom hBN were chosen to have identical thickness (10~nm). After the release of the sample onto the patterned substrate and the subsequent high temperature and vacuum annealing, electrical contacts with 5~nm Cr and 55~nm Au were fabricated by laser lithography (Heidelberg Instruments, {\micro}MLA) and electron-beam evaporation.

\textbf{Optical spectroscopy:} All optical measurements were performed at $\sim$4~K using a home-built fiber-based confocal microscope in back-scattering geometry~\cite{Neumann2017}. The samples were loaded into a closed-cycle cryostat (attocube systems, attoDRY800 or attoDRY1000) and positioned by $xyz$ piezo-stepping and $xy$ scanning units (attocube systems, ANPxyz and
ANSxy100). A low-temperature apochromatic objective with a numerical aperture of 0.81 (attocube systems, LT-APO/IR/0.81) was used to confocally excite and collect the PL signal with a focus spot size of $\sim$1~\micro\meter. A set of linear polarizers (Thorlabs, LPVIS) and half-waveplates (B. Halle, $310-1100$~nm achromatic) mounted on piezo-rotators (attocube systems, ANR240) were employed to control the polarization in excitation and detection. For steady-state PL measurements, the samples were excited by a Ti:sapphire laser (Coherent, Mira Optima 900-F) in continuous-wave mode tuned to 755~nm to match the resonance of intralayer exciton transition of MoSe$_2$ monolayer and the incident power was kept at 1~\micro\watt~unless stated otherwise. The PL signal was spectrally dispersed by a monochromator (Roper Scientific, Acton SP2500 or Acton SpectraPro 300i) with a 150 or 300 grooves/mm grating and detected by a liquid nitrogen cooled (Roper Scientific, Spec-10:100BR) or Peltier-cooled (Andor, iDus 416) charge-coupled device. For time-resolved PL measurements, a supercontinuum laser (NKT Photonics, SuperK Extreme EXW-12) with a pulse duration of 5~ps was coupled to an tunable filter (NKT Photonics, SuperK VARIA) for pulsed excitation at 755~nm with a repetition rate of 1.25~MHz. The PL signal was directed to a superconducting nanowire single-photon detector (Scontel, TCOPRS-CCR-SW-85) and the photon detection events were recorded by an electronic correlator (PicoQuant, PicoHarp 300). To prevent pile-up effects and Auger-mediated non-radiative process, the count rate of avalanche photodiode was adjusted to $< 1\%$ of the repetition rate by decreasing the incident power to 10~nW. The instrument response function (IRF) was measured by detecting the laser pulse using the same setup, and its full-width at half-maximum (FWHM) was determined to $\approx$100~ps, which is at least one order of magnitude smaller than the lifetime of interlayer excitons. It is thus unnecessary to perform deconvolution when fitting the PL decay. 


\textbf{Electrostatic control:}
For electric field measurements on the dual-gated sample (Fig.~4), the voltages were applied to the top and bottom graphene gates using two programmable DC-sourcemeters (Keithley 2400) with the MoSe$_2$ and WSe$_2$ monolayers being grounded. Since the top and bottom hBN presents the same thickness $d_\text{hBN} = 10$~nm, the out-of-plane electric field control without charge doping effects was realized by applying opposite voltages $V$ and $-V$ on the top and bottom gates, with the electric field given by $F=-V/d_\text{hBN}\times(\epsilon_\text{hBN}/\epsilon_\text{TMD})$~\cite{Jauregui2019}, where $\epsilon_\text{hBN}=3.5$ and $\epsilon_\text{TMD}=7.2$ are the relative dielectric constants of hBN and TMD, respectively~\cite{Ahmed2018,Laturia2018}.
    
\textbf{Scanning electron microscopy:} Scanning electron microscopy (SEM) imaging with secondary electron detection was conducted using a Raith eLINE system. The acceleration voltage was $\sim$2~kV and the aperture was 30~\micro\meter. For optimal secondary electron signal, samples were positioned close to the out-lens detector by an elevated sample holder, and the working distance was then reduced to $\sim$4.5~mm. In addition, samples were tilted by an angle of $\sim$38° and rotated azimuthally so that the TMD crystal axis had a angle of 10° to the bottom horizontal line of the SEM image~\cite{Rupp2023}. Under the above conditions, the \RX and \RM domains can be clearly distinguished in the samples capped by a thin top hBN (thickness $<$5~nm)~\cite{Zhao2023}.

\textbf{Torsional force microscopy:} Torsional force microscopy (TFM) imaging were performed using a commercially available AFM (Bruker, Dimension Icon equipped with a NanoScope V controller) based on the procedure developed in Ref.~\cite{Pendharkar2024}. A single crystal diamond probe with a tip radius of curvature $<5$~nm (Bruker, AD-2.8-SS) was mounted on a torsional resonance (TR) mode cantilever holder (Bruker, DTRCH-AM). By running the contact mode-like TR-dynamic friction mapping, the reconstructed moir\'e patterns were visualized in the amplitude and phase channels. To precisely determine the width of \RX stripes, the TR deflection set point was gradually stepped up, and the images were recorded once the contrast was optimized. 

\textbf{Theoretical modeling:}
Atomic reconstruction was modeled in numerical simulations by discretizing the displacement field of the HBL lattice, followed by optimization of the total energy using the L-BFGS algorithm implemented in MATLAB.
\\

\textbf{Acknowledgements:}\\
This research was funded by the Deutsche Forschungsgemeinschaft (DFG, German Research Foundation) within the Priority Programme SPP 2244 2DMP and the Germany's Excellence Strategy EXC-2111-390814868(MCQST), as well as by the Bavarian Hightech Agenda within the EQAP project. K.W. and T.T. acknowledge support from the JSPS KAKENHI (Grant Numbers 21H05233 and 23H02052) , the CREST (JPMJCR24A5), JST and World Premier International Research Center Initiative (WPI), MEXT, Japan. Z.A.I. acknowledges support of the Foundation for the Advancement of Theoretical
Physics and Mathematics ``BASIS''.

\textbf{Contributions:}\\
S.\,Z. and Z.\,Li. contributed equally to this work. 

\textbf{Corresponding authors:}\\
A.\,S.\,B. (anvar.baimuratov@lmu.de) and A.\,H. (alexander.hoegele@lmu.de). 

%
 
\end{document}


\title{Supplemental Information for \\ Control of atomic reconstruction and quasi-1D excitons \\ in strain-engineered moir\'e heterostructures} 
    
    \author{Shen~Zhao$^{1}$, Zhijie~Li$^{1,2}$, Zakhar~A.~Iakovlev$^{3}$, Peirui~Ji$^{4}$, Fanrong~Lin$^{5}$, Xin~Huang$^{6}$, Kenji~Watanabe$^{7}$, Takashi~Taniguchi$^{8}$, Mikhail~M.~Glazov$^{3}$, Anvar~S.~Baimuratov$^{1,9}$ and Alexander~H\"ogele$^{1,10}$}
    
    
    \affiliation{$^1$Fakult\"at f\"ur Physik, Munich Quantum Center, and Center for NanoScience (CeNS), Ludwig-Maximilians-Universit\"at M\"unchen, Geschwister-Scholl-Platz 1, 80539 M\"unchen, Germany}

    \affiliation{$^2$Institute of Physics, Carl von Ossietzky University, 26129 Oldenburg, Germany}
    
    \affiliation{$^3$Ioffe Institute, 194021 Saint Petersburg, Russian Federation}
    
    \affiliation{$^4$State Key Laboratory for Manufacturing Systems Engineering, Xi'an Jiaotong University, Xi'an 710049, China}
    
    \affiliation{$^5$1st Physical Institute, Faculty of Physics, University of G\"ottingen, Friedrich-Hund-Platz 1, G\"ottingen 37077, Germany}
    
    \affiliation{$^6$Key Laboratory of Nano-devices and Applications, Suzhou Institute of Nano-Tech and Nano-Bionics, Chinese Academy of Sciences (CAS), Suzhou 215123, China}
    
    \affiliation{$^7$Research Center for Electronic and Optical Materials, National Institute for Materials Science, 1-1 Namiki, Tsukuba 305-0044, Japan}
    
    \affiliation{$^8$Research Center for Materials Nanoarchitectonics, National Institute for Materials Science, 1-1 Namiki, Tsukuba 305-0044, Japan}
    
    \affiliation{$^9$Center for Engineering Physics, Skolkovo Institute of Science and Technology, Bolshoy Boulevard 30, building 1, Moscow 121205, Russia}
    
    \affiliation{$^{10}$Munich Center for Quantum Science and Technology (MCQST), Schellingstra\ss{}e 4, 80799 M\"unchen, Germany}

\maketitle

\vspace{-10pt}

\vspace{20pt}

\tableofcontents
\clearpage

\noindent \textbf{Supplementary Note $\mathbf{1}$: 1D reconstruction simulation}
\addcontentsline{toc}{section}{Note $\mathbf{1}$: 1D reconstruction simulation}

Reconstruction of MoSe\(_2\)-WSe\(_2\) heterobilayers (HBL) is driven by the interplay between interlayer adhesion and intralayer strain. The total energy of the system, which combines both contributions, can be written as an integral over the HBL area \(S\) \cite{Enaldiev2020, Carr2018}:
\begin{equation}\label{eq:totalE}
\mathcal{E} = \int_S \left[ U_1(\mathbf{r}) + U_2(\mathbf{r}) + W(\mathbf{r}) \right] \, d\mathbf{r},
\end{equation}
where \(U_1(\mathbf{r})\) and \(U_2(\mathbf{r})\) are the intralayer strain energy densities in the MoSe\(_2\) and WSe\(_2\) layers, respectively, and \(W(\mathbf{r})\) is the interlayer adhesion energy density. Both energy densities depend on the in-plane displacement fields of the individual layers.

In the case of one-dimensional domain-wall structures, we assume an initially strained system where both layers are uniformly stretched by the same strain \(\epsilon\). This corresponds to a homogeneous displacement in the \(x\)-direction, given by
\begin{equation}\label{eq:u0}
\mathbf{u}_0(\mathbf{r}) = \epsilon\, x\, \mathbf{e}_x.
\end{equation}

Assuming that the MoSe\(_2\) and WSe\(_2\) layers have identical elastic properties and lattice constants \cite{Zhao2023}, their displacement fields are related by
\begin{align}
\mathbf{u}_{\text{MoSe}_2}(\mathbf{r}) &= \mathbf{u}_0(\mathbf{r}) + \frac{\mathbf{u}(\mathbf{r})}{2} \equiv \left( \epsilon\, x + \frac{u_x(\mathbf{r})}{2}, \frac{u_y(\mathbf{r})}{2} \right), \label{eq:uMoSe2} \\
\mathbf{u}_{\text{WSe}_2}(\mathbf{r}) &= \mathbf{u}_0(\mathbf{r}) - \frac{\mathbf{u}(\mathbf{r})}{2} \equiv \left( \epsilon\, x - \frac{u_x(\mathbf{r})}{2}, -\frac{u_y(\mathbf{r})}{2} \right), \label{eq:uWSe2}
\end{align}
where \(\mathbf{u}(\mathbf{r})\) represents the relative displacement field between the two layers.

In the framework of two-dimensional linear elasticity, the intralayer strain energy density can be expressed in terms of the Lame parameters \(\lambda\) and \(\mu\). For an isotropic material, it takes the form \cite{Landau}
\begin{equation}\label{eq:Un}
U_n(\mathbf{r}) = \frac{1}{2} \lambda \left( \varepsilon_{n,xx} + \varepsilon_{n,yy} \right)^2 + \mu \left( \varepsilon_{n,xx}^2 + \varepsilon_{n,yy}^2 + 2\varepsilon_{n,xy}^2 \right),
\end{equation}
with the strain tensor components defined as
\begin{align}
\varepsilon_{n,xx}(\mathbf{r}) &= \epsilon \pm \frac{\partial u_x(\mathbf{r})}{\partial x}, \label{eq:exx} \\
\varepsilon_{n,yy}(\mathbf{r}) &= \pm \frac{\partial u_y(\mathbf{r})}{\partial y}, \label{eq:eyy} \\
\varepsilon_{n,xy}(\mathbf{r}) &= \pm \frac{1}{2} \left( \frac{\partial u_x(\mathbf{r})}{\partial y} + \frac{\partial u_y(\mathbf{r})}{\partial x} \right). \label{eq:exy}
\end{align}
Here, the index \(n = 1, 2\) corresponds to the MoSe\(_2\) and WSe\(_2\) layers, respectively, with the upper sign referring to MoSe\(_2\) and the lower sign to WSe\(_2\).

\begin{figure}[htbp]
    \centering
    \includegraphics[scale=1]{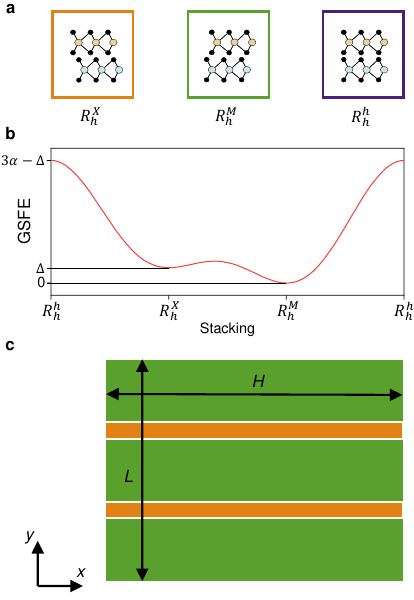}
    \caption{\textbf{a}, Schematics of high-symmetry stackings in R-type heterobilayer, with the notations adopted from Ref~\cite{Zhao2023}. Their corresponding colors are used in the Fig.~1c,d of the main text to highlighted the domain-wall structures. \textbf{b}, Generalized stacking fault energy (GSFE) function, illustrating the adhesion energy landscape as a function of the relative stacking between the layers. \textbf{c}, Schematic of stackings for the simulated of 1D reconstruction.}
    \label{fig:gsfe}
\end{figure}

The interlayer term \(W(\mathbf{r})\) accounts for the adhesion energy or generalized stacking fault energy (GSFE) between the layers and depends on the local stacking configuration as well as the relative displacement. We assume a Fourier expansion \cite{Carr2018} while significantly reducing the number of parameters:
\begin{multline}\label{eq:W}
W(\mathbf{r}) = 3\alpha - \alpha\Bigl( \cos v + \cos w + \cos(v + w) \Bigr) \\
+ \left( \frac{2\sqrt{3}\Delta}{9} - \sqrt{3}\alpha \right) \Bigl( \sin v + \sin w - \sin(v + w) \Bigr),
\end{multline}
where
\begin{equation}\label{eq:vw}
\begin{pmatrix}
v \\[1ex] w
\end{pmatrix}
=
\frac{2\pi}{a}
\begin{bmatrix}
1 & 1/\sqrt{3} \\
0 & 2/\sqrt{3}
\end{bmatrix}
\begin{pmatrix}
-\theta y + u_x(\mathbf{r}) \\
\theta x + u_y(\mathbf{r})
\end{pmatrix}.
\end{equation}
In this expression, \(\theta\) is the twist angle of the initial heterobilayer before stretching, and \(a\) is the average lattice parameter of the MoSe\(_2\) and WSe\(_2\) layers.

In Fig.~\ref{fig:gsfe}a,b we show high symmetry stackings for  MoSe\(_2\)-WSe\(_2\) HBL and the parametrization of the GFSE function according to \eqref{eq:W}. The maximum energy, given by \(3\alpha-\Delta\), corresponds to a non-optimal stacking configuration \Rh. The absolute minimum, with a GFSE value of zero, represents the most optimal stacking \RM, while a secondary optimal stacking, \RX, is characterized by a minimum with a value of \(\Delta\).

To capture the 1D reconstruction, we simplify the problem by simulating only the transverse cross-section, as depicted in Fig.~\ref{fig:gsfe}(b). This reduction allows us to solve a 1D optimization problem. Considering the symmetry of the problem, we obtain the following strain tensor components:
\begin{align}
\varepsilon_{n,xx}(\mathbf{r}) &= \epsilon, \label{eq:eps_xx} \\
\varepsilon_{n,yy}(\mathbf{r}) &= \pm \frac{1}{2}\frac{\partial u_y(\mathbf{r})}{\partial y}, \label{eq:eps_yy} \\
\varepsilon_{n,xy}(\mathbf{r}) &= \pm \frac{1}{4}\left( \frac{\partial u_x(\mathbf{r})}{\partial y} - \theta \right). \label{eq:eps_xy}
\end{align}
This simplification reduces the energy expression in Eq.~\eqref{eq:totalE} to
\begin{equation}\label{eq:E1D}
E_{1\mathrm{D}} = (\lambda + 2\mu)LH\epsilon^2 + L \int_0^H \Bigl[ (\lambda + 2\mu)\varepsilon_{1,yy}^2(y)
+ 4\mu\,\varepsilon_{1,xy}^2(y) + W(0,y) \Bigr] \, dy.
\end{equation}
Here we assume that the simulation area \(S\) is a rectangle with dimensions \(L \times H\), as shown in Fig.~\ref{fig:gsfe}c.

For the MoSe\(_2\)-WSe\(_2\) HBL, the lattice constant is \(a = 0.3288\)~nm, and the effective Lame parameters are \(\lambda = 222\)~eV/nm\(^2\) and \(\mu = 306\)~eV/nm\(^2\), \(\alpha = 60\)~meV/nm\(^2\) and \(\Delta = 1\)~meV/nm\(^2\)~\cite{Enaldiev2020,Zhao2023,Li2023}.

\begin{figure}[htbp]
    \centering
    \includegraphics[scale=1]{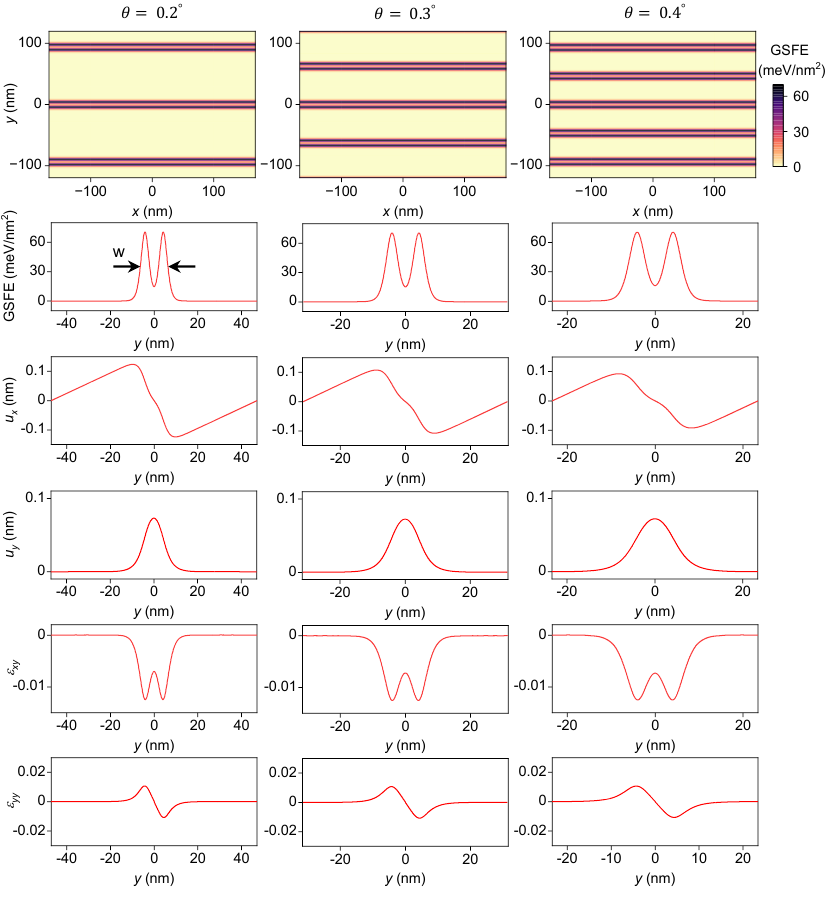}
    \caption{Simulation results for three twist angles \(\theta = 0.2^\circ, 0.3^\circ, 0.4^\circ\), shown in columns from left to right. The second row presents the cross-section of the generalized stacking fault energy function along the \(y\)-axis over one period of the 1D superlattice. The third and fourth rows show the displacement fields \(u_x\) and \(u_y\), respectively. The fifth and sixth rows display the corresponding strain tensor components \(\varepsilon_{xy}\) and \(\varepsilon_{yy}\).
}
    \label{fig:sim}
\end{figure}

We show the results of the simulation in Fig.~\ref{fig:sim}, where different columns correspond to three twist angles \(\theta = 0.2^\circ,0.3^\circ,0.4^\circ\). The second row displays the cross-section of the GSFE function along the \(y\)-axis for one period of the 1D superlattice. The third and fourth rows show the displacements \(u_x\) and \(u_y\), respectively.  Finally, the fifth and sixth rows present the strain tensor components \(\varepsilon_{xy}\) and \(\varepsilon_{yy}\).

\clearpage

\noindent \textbf{Supplementary Note $\mathbf{2}$: 1D-confined exciton fine structure}
\addcontentsline{toc}{section}{Note $\mathbf{2}$:: 1D-confined exciton fine structure}

The exciton fine structure appears as a result of the long-range exchange interaction between an electron and a hole~\cite{BP_exch71,ivchenko05a,Yu:2014fk-1,glazov2014exciton}.  The Hamiltonian describing the exchange interaction can be written in the following form~\cite{glazov2014exciton}
\begin{equation}
    \label{H:LT}
    \mathcal H_{exch}(\bm k) = \frac{\hbar\Gamma_0}{qk} 
    \begin{pmatrix}
        k_x^2 & k_x k_y\\
       k_x k_y & k_y^2
    \end{pmatrix},
\end{equation}
where the basis of linearly polarized states along the $x$ and $y$ axes is chosen, $\bm k$ is the in-plane wavevector of exciton corresponding to the center-of-mass motion, $\Gamma_0$ is the radiative decay rate of the exciton and $q { = \omega / c}$ is the wavevector of photon at the exciton resonance frequency.

We take the envelope wave function of exciton in quasi-1D confinement as in the infinite barrier approximation 
\begin{equation}
\label{psi:inf}
    \Psi(y) \propto \cos{\left( \frac{\pi y}{w}\right)},
\end{equation}
which, according to the analysis in the main text provides reasonable estimate of the exciton size-quantization energy. Combining Eqs.~\eqref{H:LT} and \eqref{psi:inf} we obtain the splitting between the $x$- and $y$-polarized excitonic states in the form 
\begin{equation}
\label{Delta:xy:inf}
    \Delta_{xy} = \frac{k_x^2-k_y^2}{k}\frac{\hbar\Gamma_0}{q} = -\frac{\pi\hbar\Gamma_0}{qw}.
\end{equation}

Equations~\eqref{H:LT} and \eqref{Delta:xy:inf} were derived assuming that a heterobilayer of negligible (compared to the $q^{-1}$) thickness is freely suspended. The calculations performed for the encapsulated structures~\cite{prazdnichnykh2020control} show that, in the general case, a screening of the long-range exchange interaction by the environment appears that can be taken into account by a weakly wavevector dependent effective dielectric constant $\varepsilon_{\rm eff}$:
\begin{equation}
    \label{Delta:xy:QW}
    \Delta_{xy} = -\frac{\pi\hbar\Gamma_0}{\varepsilon_{\rm eff}qw}.
\end{equation}
For the heterobilayer symmetrically encapsulated in hBN $\varepsilon_{\rm eff} = \varepsilon_{\rm hBN}$. In this case of a symmetric encapsulation one can use Eq.~\eqref{Delta:xy:QW} with $\Gamma_0$ and $q$ being, respectively, the radiative decay rate of exciton into hBN and the wavevector of light the hBN. If the cap layer is thin enough that $d_{\rm cap} \ll a$ then $\varepsilon_{\rm eff} = (1+\varepsilon_{\rm hBN})/2$. 

For rough estimates we take $\Gamma_0^{-1} = 50$~ps that corresponds to a radiative decay of interlayer exciton within the light cone~\cite{Barr2022,Wietek2024}, $\hbar\omega = E_X = 1.34$~eV (main text), $a = 3$~nm (minimal $w = 3$~nm from main text), $\varepsilon_{\rm eff} = 3.5$ (hBN, main text, methods):
\begin{equation}
    |\Delta E| = \frac{\pi\hbar\Gamma_0}{\varepsilon_{\rm eff} qw} = 2~\text{meV}.
\end{equation}
This value is in reasonable agreement with experiment.
\vspace*{5cm} 
\begin{figure*}[h!]
	\hspace*{-3cm} 
    \includegraphics[scale=1.3]{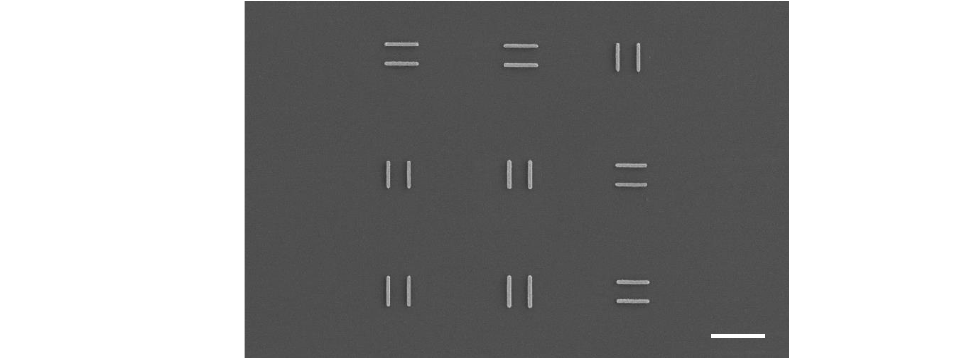}      
    \caption{SEM image showing the 3$\times$3 mesa pair pattern. The scale bar is 3~\micro\meter.}
    \label{monolayer}
\end{figure*}

\clearpage

\begin{figure*}[t!]
    \includegraphics[scale=1]{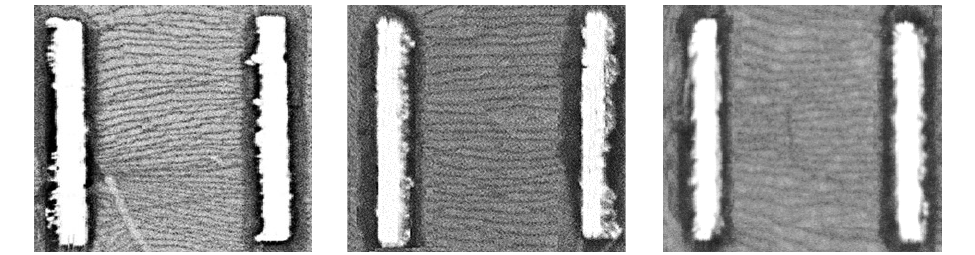}
    \caption{Electron channeling contrast SEM images of mesa pair regions with 1D arrays.}
    \label{monolayer}
\end{figure*}

\begin{figure*}[t!]
    \includegraphics[scale=1]{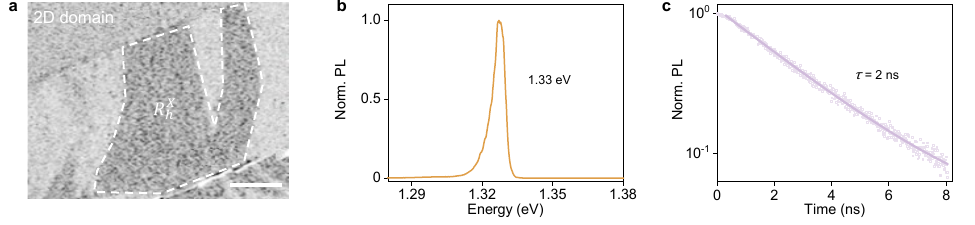}
    \caption{\textbf{a}, Electron channeling contrast SEM image showing a micron-sized 2D domain of \RX stacking (delimited by the dashed line). \textbf{b,c}, Spectrum (\textbf{b}) and decay trace (\textbf{c}) of interlayer excitons PL from the 2D domain in \textbf{a}. From these results, we obtain the energy position of free IX in \RX stacking as 1.33~eV and their lifetime of 2~ns. The lifetime is obtained from a single exponential decay fit (line in \textbf{c}), in consistent with literature~\cite{Wietek2024}.}
    \label{monolayer}
\end{figure*}

\begin{figure*}[t!]
	\hspace*{-3cm}    
    \includegraphics[scale=1.3]{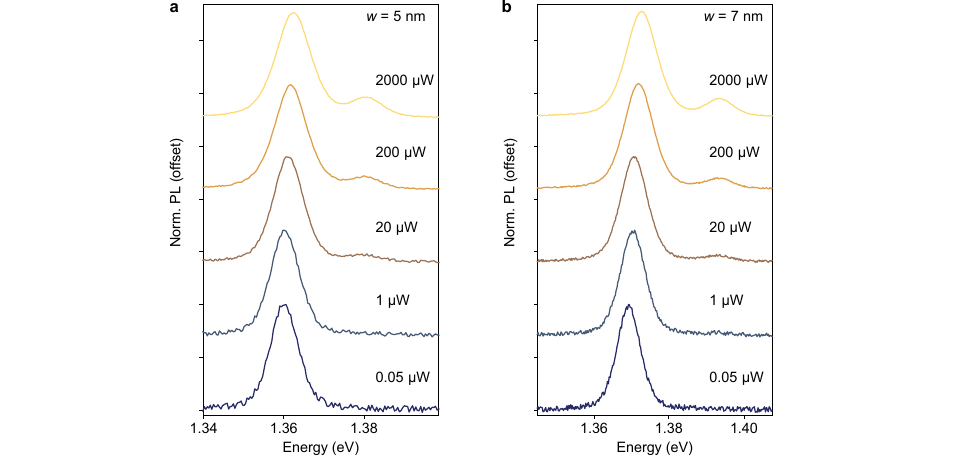}
    \caption{Power-dependent PL spectra of 1D-confined interlayer excitons for the domain wall width $w$ of 5~nm (\textbf{a}) and 7~nm (\textbf{b}). In both cases, a secondary peak emerges approximately 20~meV above the primary emission as the excitation power increases.}
    \label{monolayer}
\end{figure*}

\begin{figure*}[t!]
	\hspace*{-3cm}
    \includegraphics[scale=1.3]{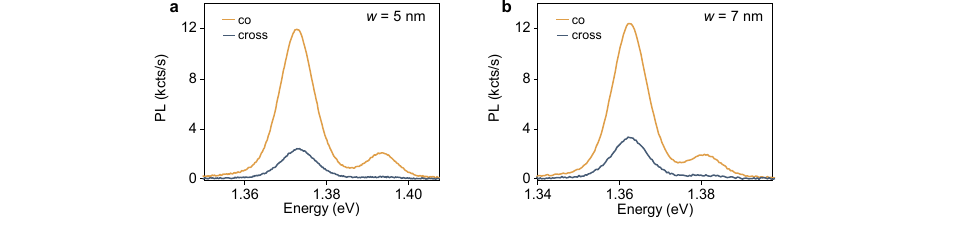}
    \caption{Linear polarization-resolved spectra of 1D-confined interlayer exciton PL for $w$ of 5~nm (\textbf{a}) and 7~nm (\textbf{b}) at an elevated excitation power of 200~\micro\watt. Akin to the primary peak, the second peak features linear polarization along the domain wall axis.}
    \label{monolayer}
\end{figure*}


%